\documentclass[printer]{aa}   

\usepackage{graphicx,url,twoopt,natbib}
\usepackage[varg]{txfonts}           
\usepackage{pdfcomment,acronym}      
\usepackage{natbib}
\usepackage{graphics}
\usepackage{longtable}

\usepackage{latexsym}
\usepackage[T1]{fontenc}       
\usepackage{bm}
\usepackage{amsmath,amsfonts,amssymb}
\usepackage{datetime2}

\begin{document}


\title{Thermal desorption of formamide and methylamine from graphite and amorphous water ice surfaces}

\titlerunning{Thermal desorption of formamide and methylamine}
\authorrunning{H. Chaabouni et al.}

\author {H. Chaabouni
          \and S. Diana
          \and T. Nguyen
          \and F. Dulieu }

\institute{LERMA, Universit\'{e} de Cergy-Pontoise, Observatoire de Paris, PSL Research University, Sorbonne Universit\'{e}, UPMC Univ. Paris 06, UMR 8112, CNRS,
5 mail Gay Lussac, 95000, Cergy-Pontoise, France. \\
\email{Henda.Chaabouni@u-cergy.fr}\\ \email{Francois.Dulieu@obspm.fr} \label{inst1}}

\date{Received 18 April 2017/ Accepted 27 December 2017}

\abstract{Formamide (${\rm NH_2CHO}$) and methylamine (${\rm CH_3NH_2}$)  are known to be the most abundant amine-containing molecules in many astrophysical environments. The presence
of these molecules in the gas phase may result from thermal desorption of interstellar ices.} {The aim of this work is to determine the values of the desorption energies of formamide and methylamine from analogues of interstellar dust grain surfaces and to understand their interaction with water ice.} {Temperature programmed desorption (TPD) experiments of formamide and methylamine ices were performed in the sub-monolayer and monolayer regimes on graphite (HOPG) and non-porous amorphous solid water (np-ASW) ice surfaces at temperatures 40-240~K. The desorption energy distributions of these two molecules were calculated from TPD measurements using a set of independent Polanyi-Wigner equations.}
{The maximum of the desorption of formamide from both graphite and ASW ice surfaces occurs at 176~K after the desorption of ${\rm H_2O}$ molecules, whereas the desorption profile of methylamine depends strongly on the substrate. Solid methylamine starts to desorb below 100~K from the graphite surface. Its desorption from the water ice surface occurs after 120~K and stops during the water ice sublimation around 150~K. It continues to desorb from the graphite surface at temperatures higher than160~K. }
{More than 95~\% of solid ${\rm NH_2CHO}$ diffuses through the np-ASW ice surface towards the graphitic substrate and is released into the gas phase with a desorption energy distribution  ${\rm E_{des}=7460-9380~K}$, which is measured with the best-fit pre-exponential factor A=10$^{18}$~s$^{-1}$. However, the desorption energy distribution of methylamine from the np-ASW ice surface (${\rm E_{des}}$=3850-8420~K) is measured with the best-fit pre-exponential factor A=10$^{12}$~s$^{-1}$. A fraction of solid methylamine monolayer of roughly 0.15 diffuses through the water ice surface towards the HOPG substrate. This small amount of methylamine desorbs later with higher binding energies (5050-8420~K) that exceed that of the crystalline water ice (${\rm E_{des}=4930 ~K}$), which  is calculated with the same pre-exponential factor A=10$^{12}$~s$^{-1}$. The best wetting ability of methylamine compared to ${\rm H_2O}$ molecules makes ${\rm CH_3NH_2}$ molecules a refractory species for low coverage. Other binding energies of astrophysical relevant molecules are gathered and compared, but we could not link the chemical functional groups (amino, methyl, hydroxyl, and carbonyl) with the binding energy properties. Implications of these high binding energies are discussed.}


\maketitle

\section{Introduction}

Formamide, ${\rm NH_2CHO}$, is a species of great relevance in prebiotic chemistry \citep{2012Saladino,2015Barone}. This species contains a (quasi) peptide bond (-NH-(C=O))
that is very active in the synthesis of nucleic acid bases, carboxylic acids, sugars \citep{2012Saladino}, and amino acids responsible for the formation of proteins. It is also well known to be the precursor species for the  formation of purine and pyrimidine bases during the course of chemical evolution leading to the origin of life \citep{2016Bhushan}.
Formamide  has been observed in the gas phase in several astronomical environments, such as prestellar and protostellar objects \citep{2013Kahane}, massive hot molecular cores
\citep{2007Bisschopp}, hot corinos \citep{2015Sepulcre}, and comets, such as C/20012 F6 (Lemmon), C/2013 R1 (Lovejoy)  \citep{2014biver}, and Hale-Bopp \citep{2000bockelee}.

Formamide has been once tentatively identified in interstellar ices of star-forming regions by ISO-SWS infrared spectra \citep{2004Raunier}, but it is no longer in the list of confirmed molecules in interstellar ices \citep{2015Boogert}.
As for numerous molecules, especially interstellar complex organic molecules, astrochemists would like to put constraint on the formation pathway of formamide. In order to discriminate different scenarios of star formation, solid-state and gas-phase formation routes are usually debated because they imply different physical conditions. For both chemical routes, there are arguments coming from observations and others belonging to the physical or chemical properties of ${\rm NH_2CHO}$.
On the observational side, \cite{2015Sepulcre} have noticed that  ${\rm NH_2CHO}$ is correlated to isocyanic acid (HNCO) abundances. Furthermore, the first detection of deuterated formamide towards the low protostar IRAS 16293 by ALMA-PILS, which was a D/H ratio of 2~\% similar to HNCO \citep{2016Coutens}, has been found to be in agreement with the hypotheses that these species are chemically related through grain-surface formation.
On the other hand, \cite{2017Codella} have explained well the observation of formamide in the brightest shocked region of the large-scale molecular outflow L1157-B1 through the gas-phase reactions between ${\rm NH_2}$ and ${\rm H_2CO}$.
For prestellar core conditions ($T<10~K$), despite the inclusion of this very efficient reaction in their chemical network, \cite{2017Vasyunin} do not really overestimate the abundance of formamide in the prestellar core L1544.  In these cold prestellar sources, formamide remains undetected according to \cite{2016Jimenez}. The scenario of the gas-phase formation route of formamide relies on the computational work of \cite{2015Barone} and that of \cite{2017skouteris}, who studied the deuteration aspects and confirmed that the gas-phase route for the formation of formamide in the first cold prestellar phase is in agreement with the observations. The theoretical discussion \citep{2016Song} about the presence or the absence of the barrier for this key reaction (${\rm NH_2 + H_2CO~\rightarrow~NH_2CHO + H}$) in the gas phase will end when experimental work is carried on.
On the other hand, in the solid state, many experimental reports about the formation of  formamide have been published.
Energetic electron bombardment of ${\rm CO-NH_3}$ ice mixtures \citep{2011Jones}, ion irradiation of ${\rm H_2O-HCN}$ ices at 18~K \citep{2004Gerakines}, and
during the warm-up of photolyzed ice mixtures of ${\rm H_2O}$, ${\rm CH_3OH}$, CO, and ${\rm NH_3}$ \citep{1995Bernstein}, led to the formation of ${\rm NH_2CHO}$. Recent laboratory experiments of \cite{2016Fedoseev} have also produced formamide molecules by hydrogenation and UV photolysis of NO in CO-rich interstellar ice analogues.
The chemical link in the solid phase of HNCO and ${\rm NH_2CHO}$ has been ruled out by H-bombardment experiments of HNCO at low surface temperature \citep{2015Noble}, which do not lead to detectable amounts of ${\rm NH_2CHO}$ molecules. Our group demonstrates that formamide can be produced with high efficiencies even without the help of external energy (Nguyen et al in preparation).
A last aspect of the solid phase chemical route concerns the binding energy of formamide.  \cite{2014Dawley} have investigated thermal desorption of thick mixed ${\rm H_2O-NH_2CHO}$ ices on a silicate (${\rm SiO_2}$) grain analogue from 70~K to 400~K. These authors have observed a delayed desorption peak of the water ice at 160-200~K, resulting from the diffusion of
${\rm H_2O}$ molecules during the phase transition of formamide accumulated in the ice, followed by a large desorption profile at higher temperatures (200-380~K), corresponding to the diffusion of molecules through the high surface area of silicate. The desorption activation energy of  formamide measured experimentally from ${\rm SiO_2}$ substrate has been found to be ${\rm 14.7~kcal\cdot mol^{-1}}$ (or 7397~K) with a pre-factor ${\rm A=~10^{13}~s^{-1}}$. Recently, \cite{2017Wakelam} have estimated the binding energy of formamide on water ice surface to be  ($6300 \pm 1890$)~K using a semi-empirical  theoretical approach. The discrepancy can obviously come from the calculation method, but  could also find its origin in the amount of ${\rm NH_2CHO}$ interacting with the water ice, which is high in the first case \citep{2014Dawley} and very low in the second case \citep{2017Wakelam}. Experiments performed with thin layers on the surface of the substrate are required for a better comparison. Anyhow, formamide has a very high value of binding energy that is much higher than that of ${\rm H_2O}$, which is around 4930~K for crystalline water ice, estimated from the results of \cite{2001Fraser} using a pre-exponential factor ${\rm A=~10^{12}~s^{-1}}$.
At the current state of the astrochemistry,  it is not mandatory to involve the direct formation of formamide on dust grains to explain its observation (or non-observation) in various media, as long as the gas-phase reaction between ${\rm NH_2}$ and ${\rm H_2CO}$ is supposed to be efficient. If the correlation between HNCO and ${\rm NH_2CHO}$ is confirmed at lower spacial scales, it is not due to an obvious link with the solid phase. Finally,  even if  formamide is easily formed on grains, its return into the gas phase is often more puzzling than other complex organic molecules because of its high binding energy and its very low interaction with water. In summary, the main formation route of ${\rm NH_2CHO}$, by gas-phase or surface reactions, has not yet been settled and more observational and experimental work is needed.
Another interesting fundamental organic compound in biochemistry studied in this work is the methylamine, ${\rm CH_3NH_2}$, which is known to be the most abundant amine-containing molecule after formamide (${\rm NH_2CHO}$). Methylamine has been detected in the gas phase through its ${\rm 2_{02}->1_{10} A_{a^{-}}}$ state transition in the direction of Sgr B2 and Ori A~\citep{1974Fourikis} towards the giant molecular cloud Sgr B2(N) with a fractional abundance as high as ${\rm 3\times10^{-7}}$ relative to molecular hydrogen~\citep{2000Nummelin}. It has also been detected in the coma of comet 67P/Churyumov-Gerasimenko \citep{2016Altwegg} by the Rosetta space mission. In the gas phase, the formation of the methylamine molecules is completely dependent on radicals ${\rm CH_3^{\bullet}}$ and ${\rm NH_2^{\bullet}}$ produced by UV photons \citep{2008Garrod}. In the solid phase, the hydrogenation experiments of HCN molecules by H atoms at low surface temperature \citep{2011Theule} have shown the formation of the fully saturated species ${\rm CH_3NH_2}$ with methanimine as an intermediate. The desorption of these molecules occurred between 140~K and 150~K, at slightly lower temperatures than the desorption of the amorphous water ice in their experimental conditions.
Recent thermal desorption experiments of \cite{2016Souda} have shown that methylamine-water interaction is influenced by the porosity of the ASW ice. An efficient incorporation of
${\rm CH_3NH_2}$ molecules in the porous water ice film, deposited on Nickel Ni (111) substrate at 20~K has been observed, with a narrow desorption peak at 160~K, during water ice crystallization. In contrast, no diffusion of ${\rm CH_3NH_2}$ molecules in the film interior of the non-porous water ice, deposited at surface temperature 120~K, has been observed in its TPD spectra, even by hydrogen bonding formation with ${\rm H_2O}$ molecules. Despite this advanced work, no experimental values for the desorption energies of methylamine have been provided on water ice. However, a desorption energy value (6584~K) of ${\rm CH_3NH_2}$ is given in (http://kida.obs.u-bordeaux1.fr/species/154/CH3NH2.html) from a previous estimation of the OSU gas-grain code of Eric Herbst's group.
Within molecular clouds, the desorption of molecules from icy mantles back into the gas-phase occurs either by thermal, or non-thermal processes.  Non-thermal desorption of molecules may result  from the exothermic reactions occurring on the grain surface \citep{1993Duley, 2013Vasyunin, 2013Dulieu, 2016Minissale} and the impact of cosmic rays or ultraviolet photons \citep{1985leger,1990Hartquist}. Thermal desorption process occurs during the collapse of the dense clouds and the birth of protostars \citep{2007Garrod}. The gravitational energy is converted into radiation, provoking the warm-up of the grain and the desorption of molecules from the icy mantle to the gas environment. In laboratory, temperature-programmed desorption (TPD) is the most effective method for the measurement of ice sublimations over a significantly shorter timescale and for the determination of the energy required for desorption.

In this work, we investigate thermal desorption experiments of formamide (${\rm NH_2CHO}$) and methylamine (${\rm CH_3NH_2}$) adsorbate  in the sub-monolayer and monolayer regimes on two surfaces. These surfaces are that of the highly orientated pyrolytic graphite (HOPG) as a laboratory model for carbonaceous grains and the surface of the ${\rm H_2O}$ non-porous amorphous solid water (np-ASW) ice, covering the HOPG substrate and acting as an extremely relevant astrophysical surface analogue.
The aim of this work is to understand the interaction of these two amino molecules with the  water ice  and to study the effect of the graphitic substrate on the desorption processes. For this purpose, we determine the desorption energy distributions of formamide and methylamine on both surfaces, using TPD measurements and Polanyi- Wigner equation. We compare these distributions to the binding energy of pure water ice.
This paper is organized as follows: Section~\ref{exp} explains the experimental methods, Section~\ref{results} presents the experimental results and the TPD measurements of formamide (${\rm NH_2CHO}$) and methylamine (${\rm CH_3NH_2}$) on graphite (HOPG) and on np-ASW ice surfaces, Section~\ref{Analysis} describes the model developed to measure the desorption
energy distributions  of formamide and methylamine molecules from graphite and water ice surfaces,  Section~\ref{Discussion} presents the discussion of the results, and finally Section~\ref{conclusions} summarizes the main conclusions of this work.

\section{Experimental methods}\label{exp}

The experiments were performed using the FORMOLISM (FORmation of MOLecules in the InterStellar Medium) apparatus. The set-up is dedicated to study the interaction of atoms and molecules on
surfaces of astrophysical interest. The experimental set-up is briefly described here and more details are given in earlier papers \citep{2007Amiaud, 2013Chaabouni}. The apparatus is composed
of an ultra-high vacuum (UHV) stainless steel chamber with a base pressure $10^{-10}$~mbar. The sample holder is located in the centre of the main chamber. It is thermally connected to a cold finger of a closed-cycle He cryostat. The temperature of the sample is measured in the range 40~-~350~K by a calibrated platinum (Pt) diode clamped to the sample holder, which  is made of
1~cm diameter copper block, covered with a highly orientated pyrolytic graphite (HOPG, ZYA-grade) slab.

In our laboratory experiments, the viscous, liquid formamide is introduced in a pyrex flask, which is placed in a ceramic bath containing a silicone oil acting as a regulator for the temperature. Because of the low source vapour pressure of formamide in the beam line at room temperature, the liquid is warmed up to 50~$^{\circ}$C using a heating plate with a motor speed of 280 revolutions per minute. Methylamine compound is also used  under its aqueous solution with a weight percent in water of 40 Wt \%, meaning that there is 40 g of ${\rm NH_2CH_3}$ solute  for every 100 g of water solvent. Methylamine is more volatile than formamide and its source pressure at the inlet of  the beam line is set to 1.00~mbar at room temperature. At these experimental conditions, formamide and methylamine beams were aimed at the surface of the HOPG substrate, held at 40~K, for several deposition times, using a triply differentially pumped beam line, orientated 60$^{\circ}$ with respect to the surface of the sample holder.

The UHV chamber is also equipped with a movable quadrupole mass spectrometer (Hiden Analytical QMS) operating in the range 1-100 amu (atomic mass unit) with a Channeltron detector. The QMS is orientated face-on to the beam line to characterize the gas composition of formamide and methylamine beams. It is placed 5~mm in front of the surface of the sample holder to apply the temperature programmed desorption (TPD) technique using a custom LabView software that maintains a linear heating rate of ${\rm 0.2~K\cdot s^{-1}}$. During the warm-up phase of the sample, from ${\rm 40~K}$ to ${\rm 240~K}$, the species desorbing from the surface into the gas phase are ionized, or fragmented into cracking patterns, by electron impact in the ion source of the QMS. The TPD curves shown in next sections exhibit the most intensive signals monitoring with the QMS of the intact parent molecules ${\rm NH_2CHO}$ (m/z=45~amu),  ${\rm NH_2CH_3}$ (m/z=31 amu), and ${\rm H_2O}$ (m/z=18 amu), which are only converted into positive ions by electron impact ionization. The QMS is also moved to the upper position to prepare the films of the non-porous amorphous solid water (np-ASW) ice  on the graphite surface, using a micro-channel array doser (1~cm in diameter), located 2~cm in front of the surface in the UHV chamber.
The np-ASW ice film of 10~ML thickness  is grown on top of the graphite (HOPG) surface, maintained at the temperature of 110~K, by spraying water vapour under a constant deposition pressure in the vacuum chamber of ${\rm 2\times10^{-9}}$~mbar. Water vapour is obtained from deionized liquid ${\rm H_2O}$, purified by several pumping cycles under cryogenic vacuum conditions~\citep{2012Noble}. After deposition, the temperature of the surface is kept constant at 110~K for 30~minutes, until the background pressure in the vacuum chamber is stabilized at ${\rm 10^{-10}}$~mbar. Then the sample was cooled  to the base temperature 40~K prior to formamide or methylamine deposition. During the cooling of the water ice from 110  to 40~K, the structure and the morphology of the water ice is not changed. Water ice deposited at 110~K  remains in the same compact (non-porous) amorphous state at 40~K.

Since our desorption experiments are performed in the sub-monolayer and monolayer regimes, the dimensionless surface coverage (${\rm \theta}$=${\rm N/N_{mono}\leq 1}$) is defined as the number density of molecules exposed on the surface (N) by  the maximum surface number density of molecules ${\rm N_{mono}}$ prior to  multi-layer desorption \citep{2012Noble}. At surface coverage saturation ($\theta$=1), the first monolayer of molecules covering the surface is defined as the maximum number density of molecules that populate 10$^{15}$ adsorption site per cm$^{2}$ on a flat surface, such as graphite. This definition of one monolayer unit (1~ML= 10$^{15}$~${\rm molecules\cdot cm^{-2}}$) is kept over this study; this includes amorphous surfaces, such as non-porous ASW ice, where the saturation of the first layer of molecules can thus be observed at a value, which differs from the exact value of the defined 1~ML.
In our experiments, the first monolayer of solid formamide  covering the surface of graphite at 40~K\ was reached after 90 minutes of exposure time, and that of solid methylamine after beam deposition during 10 minutes. The fluxes of formamide and methylamine molecules coming from the gas phase and hitting the surface of the sample holder are defined as the ratio of the amount of molecules in the designed 1~ML divided by the  exposure time required for surface saturation. These fluxes derived from TPD data and the method described in \citep{2012Noble} are estimated to be 2.6$\times$ 10$^{10}$~${\rm molecules \cdot cm^{-2} \cdot s^{-1}}$ for ${\rm NH_2CHO}$ beam and 1.6$\times$ 10$^{12}$~${\rm molecules \cdot cm^{-2}\cdot s^{-1}}$ for ${\rm CH_3NH_2}$ beam.

\section{Experimental results}\label{results}

\subsection{Formamide} \label{Formamide}

\subsubsection{Formamide on graphite surface} \label{X}

\begin{figure}
\centering
\includegraphics[width=8.8cm]{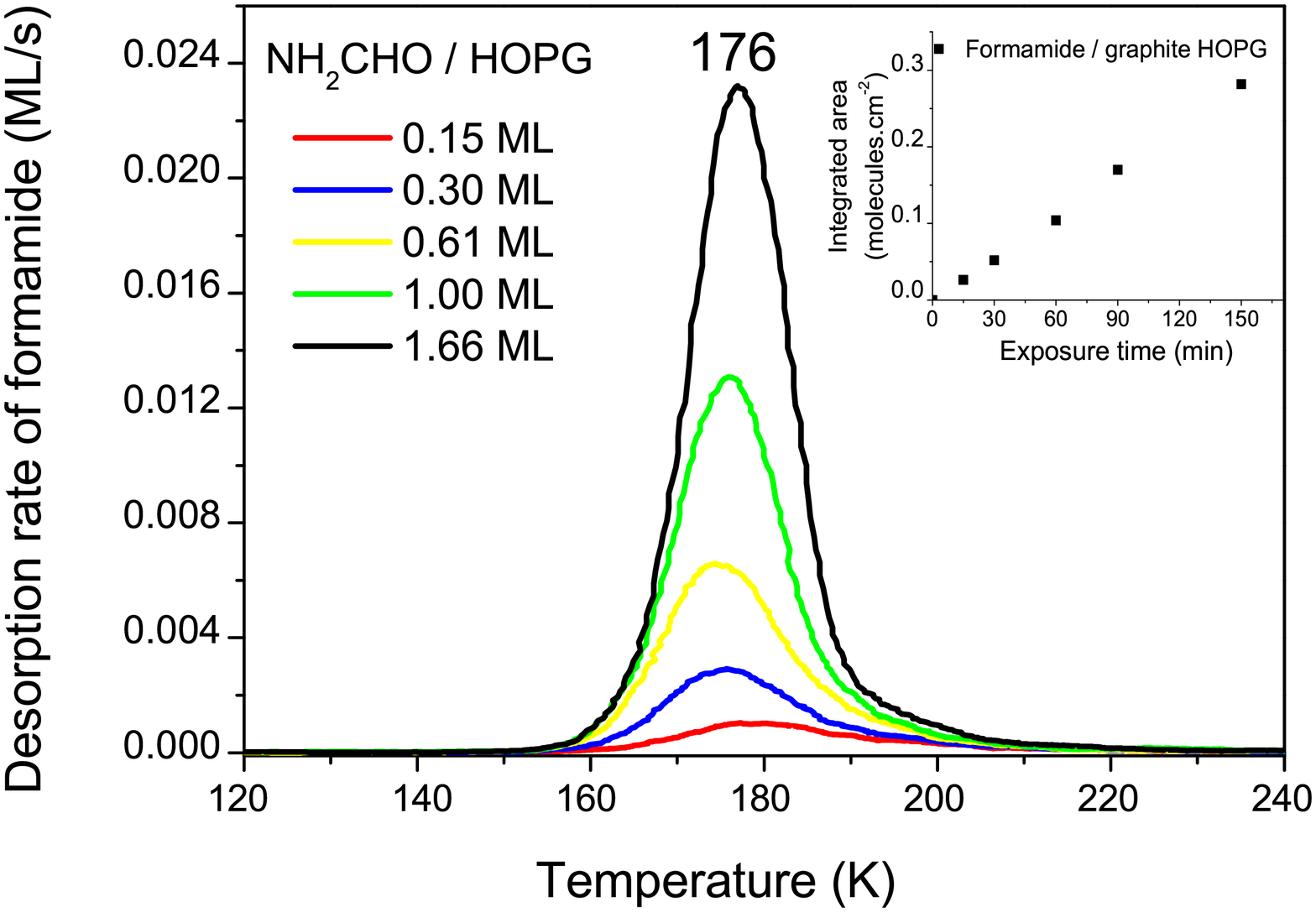}
\caption{Desorption rates of formamide (${\rm NH_2CHO}$; m/z=45), expressed in (${\rm ML\cdot s^{-1}}$) as a function of the temperature of the surface for several exposure doses
(0.15~ML, 0.30~ML, 0.61~ML, 1.00~ML, and 1.66~ML) of solid formamide on the cold HOPG surface at 40~K. The linear heating rate of the samples is ${\rm 0.2~K \cdot s^{-1}}$. The linear dependence of the integrated area under the TPD curves as a function of the exposure time is shown in the inset.} \label{Fig1}
\end{figure}

Figure~\ref{Fig1} shows the desorption curves of formamide (${\rm NH_2COH}$) given by the QMS signals of m/z=45, for several exposure doses (0.15~ML, 0.30~ML, 0.61~ML, and 1.00~ML) on the
graphite HOPG surface. The fifth TPD spectrum corresponds to the multi-layer regime of formamide given by the higher exposure dose of 1.66~ML. All the TPD spectra exhibit desorption peaks
between 160~K and 200~K. Although the integrated area under the TPD curves increases with the exposure dose of formamide  from 0.15~ML to 1.66~ML,  the
desorption peak remains at the same temperature of 176~K, mainly for surface coverage higher than 0.30~ML. For the lowest dose of 0.15 ML, the TPD peak is shifted to a slightly higher temperature compared to the higher doses.The inset in Figure~\ref{Fig1} shows the linear evolution of the integrated areas under the TPD curves (for surface dose up to 1~ML
and above) as a function of the exposure times.

\subsubsection{Formamide on np-ASW ice}\label{Formamaide water}

\begin{figure}
\centering
\includegraphics[width=8.8cm]{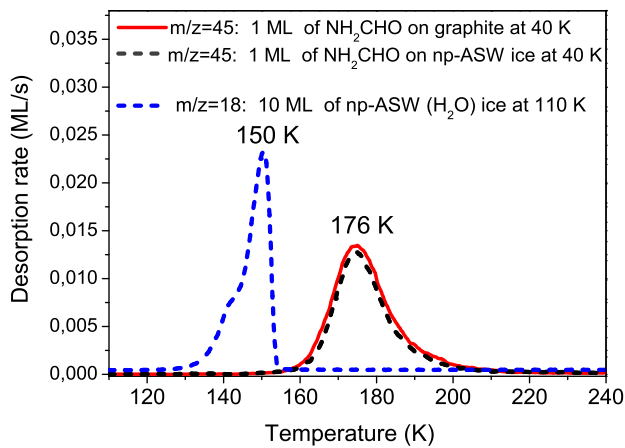}
\caption{TPD curves of formamide (${\rm NH_2CHO}$) giving the desorption rates in (${\rm ML\cdot s^{-1}}$) of the mass (m/z=45) as a function of the temperature of the surface in (K). The red line indicates the TPD signal of ${\rm NH_2CHO}$ deposited on graphite (HOPG) surface at 40~K; the black dashed line shows the TPD signal of ${\rm NH_2CHO}$ deposited on  the np-ASW ice surface at 40~K;  and the blue dashed line indicates the TPD signal of ${\rm H_2O}$ (m/z=18) for 10 ML thickness of the np-ASW ice film  prepared at 110~K on the HOPG surface and cooled down to 40~K (scaled by 0.0007).} \label{Fig2}
\end{figure}

Figure~\ref{Fig2} shows the desorption curve of solid ${\rm NH_2CHO}$ (m/z=45) on np-ASW ice, peaking at 176~K. For comparison, the desorption curve of the non-porous ASW
ice film, given by the signal  (m/z=18)  at 150~K, is also shown. Formamide is clearly desorbing from the surface at a higher temperature than the water ice. The TPD curve of formamide from the water ice surface is similar to that obtained previously from the graphite HOPG surface for the same exposure dose of 1~ML. This means that the desorption of formamide molecules into the gas
phase is not affected by the water ice substrate. The calculations of the integrated areas below the TPD curves show that more than 0.95~ML of solid formamide is likely to diffuse from the surface of the water ice towards
the graphite HOPG substrate.This diffusion probably occurs during the warm-up phase of the sample and the reorganization of the water ice. Since our TPD experiments last  few minutes, the diffusion timescale of the
molecules from the surface of the water ice to the graphitic substrate is expected to be about few seconds. Because formamide desorbs after water, its binding energy from the HOPG surface is expected to be higher than that of ${\rm H_2O}$ molecules.

\subsection{Methylamine}\label{methylamine}

Similarly to formamide, the adsorption-desorption experiments of methylamine ${\rm CH_3NH_2}$ molecules were investigated both on graphite (HOPG) and on np-ASW ice surfaces.

\subsubsection{Methylamine on graphite surface} \label{Z}

\begin{figure}
\centering
\includegraphics[width=8.8cm]{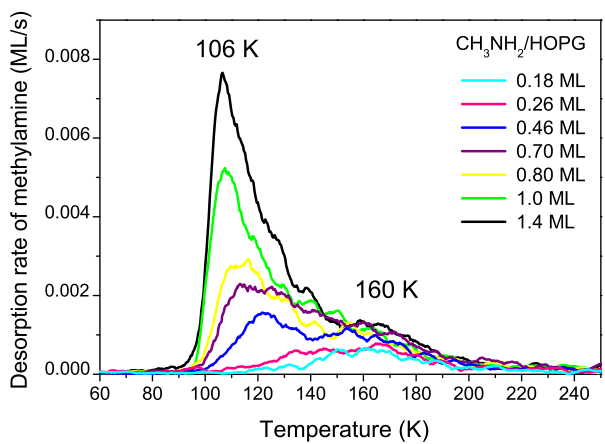}
\caption{Desorption rates, expressed in (${\rm ML\cdot s^{-1}}$), of methylamine (${\rm CH_3NH_2}$)  as a function of the temperature of the surface in (K) for several exposure doses of solid methylamine
(0.18~ML, 0.26~ML, 0.46~ML, 0.70 ML, 0.80 ML, 1.0~ML, and 1.40~ML) on the cold HOPG surface at 40~K. The linear heating rate of the samples is ${\rm 0.2~K \cdot s^{-1}}$.} \label{Fig3}
\end{figure}

Figure~\ref{Fig3} shows the desorption curves of methylamine (m/z=31) for several exposure doses in (ML) on graphite surface held at 40~K. For very small doses (0.18 ML, 0.26 ML), molecules
occupy the most energetically favourable adsorption sites and are tightly bound to the surface of the HOPG, desorbing late from the surface, at temperatures up to 160~K. This small TPD peak  is
assigned to the strong interaction of the ${\rm NH_2CH_3}$ molecules with the graphitic surface in the sub-monolayer regime. As the surface coverage increases, molecules are forced to populate
progressively less tightly bound sites of the graphitic HOPG surface, provoking earlier desorption at about 113~K. The area of the second desorption peak of ${\rm CH_3NH_2}$ at 113~K increases
with the increase of the exposure doses in the sub-monolayer regime, and its maximum shifts towards a lower temperature (106~K) once the surface coverage is saturated and the first monolayer is formed. The strong desorption peak at 106~K for 1.40 ML corresponds to the onset of the multi-layer desorption of solid methylamine, where ${\rm CH_3NH_2}$ is bound to ${\rm CH_3NH_2}$
ice by hydrogen bonds.

\subsubsection{Methylamine on np-ASW ice} \label{Methylamine on ASW ice}

\begin{figure}
\centering
\includegraphics[width=8.8cm]{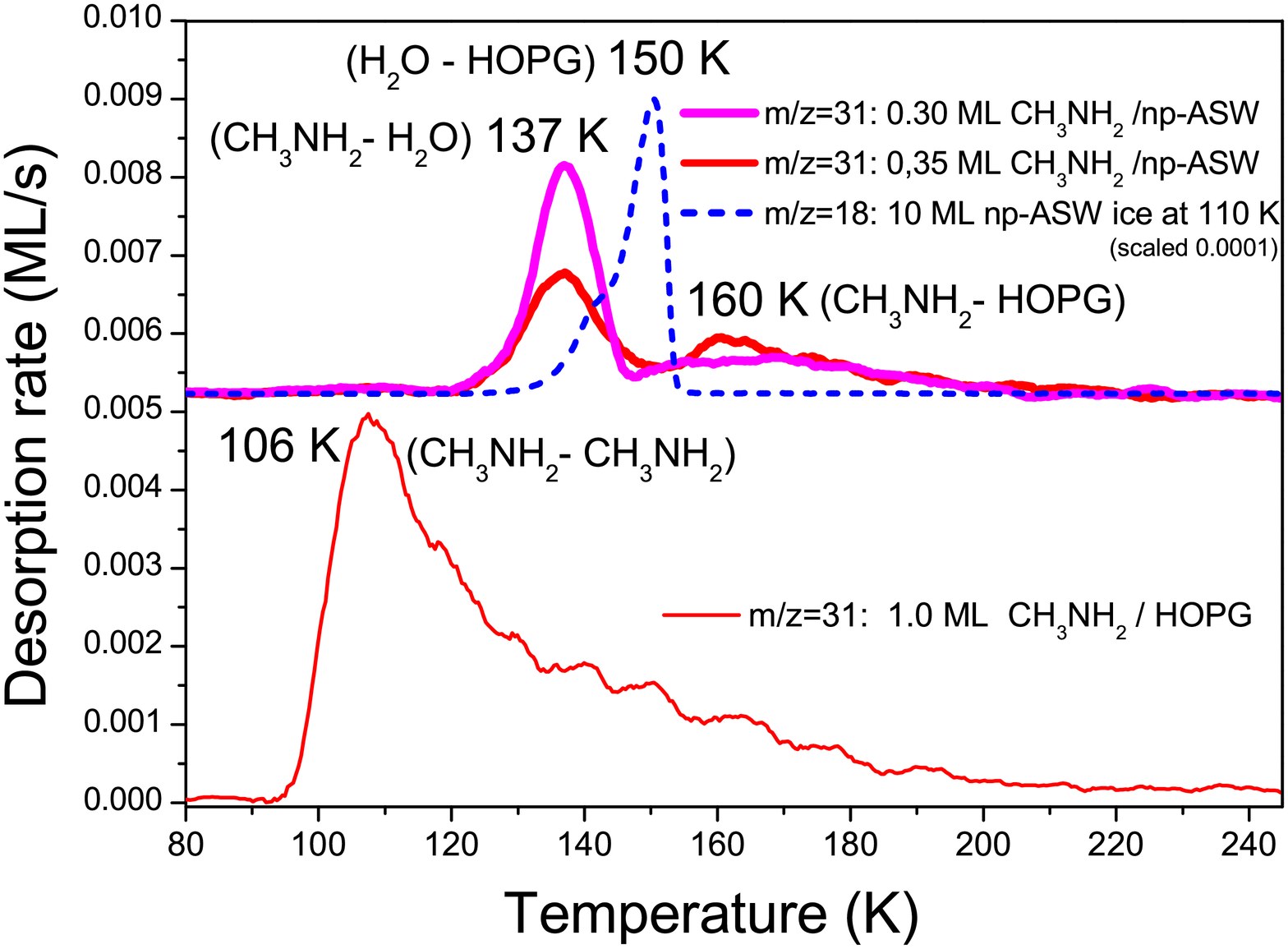}
\caption{TPD curves of methylamine giving the desorption rates in (${\rm ML\cdot s^{-1}}$) of the mass (m/z=31) as a function of the temperature of the surface in (K). The thin red line indicates 1~ML of ${\rm CH_3NH_2}$  deposited on graphite HOPG substrate at 40~K; the thick red line represents 0.30~ML of ${\rm CH_3NH_2}$  deposited on the np-ASW ice surface held at 40~K; the thick magenta line indicates 0.35~ML of ${\rm CH_3NH_2}$ deposited on the np-ASW ice surface at 40~K; and the blue dashed line represents TPD curve of ${\rm H_2O}$ (m/z=18) for 10~ML of the np-ASW ice film  prepared at 110~K on the HOPG surface and cooled down to 40~K (scaled by 0.001).} \label{Fig4}
\end{figure}

Figure~\ref{Fig4} shows the TPD curves of solid methylamine (m/z=31) from the np-ASW ice surface for two small doses (0.30~ML and 0.35~ML) deposited on the surface during 5 and 6
minutes, respectively (thick red and magenta lines). These curves are compared to the TPD curve of solid methylamine  on the graphite (HOPG) surface for a 1~ML exposure dose (thin red line) and to the  ${\rm H_2O}$ TPD curve
(m/z=18) of pure np-ASW ice, with 10~ML thickness, grown on the HOPG surface at 110~K (blue dashed line).
As shown in Figure~\ref{Fig4}, the large TPD peak of solid ${\rm CH_3NH_2}$ (1~ML) from the HOPG surface shows a multi-layer desorption peak at lower surface temperatures centred at 106~K, a shoulder at about 140-150~K resulting from the mixture of the methylamine with the water coming from the aqueous phase, and a long desorption tail at higher temperatures, up to 220~K.
When ${\rm CH_3NH_2}$ is deposited on top of the np-ASW ice film of 10~ML thickness, prepared at 110~K and cooled down to 40~K, the maximum of the desorption peak of methylamine is shifted to the higher temperature, 137~K, and a large desorption peak with a maximum centred at 160~K appears as a tail at higher surface temperatures up to 220~K. This desorption behaviour of methylamine on the np-ASW ice is observed for two close exposure doses of 0.30~ML and 0.35~ML.
The desorption peaks at 137~K, which occur before the desorption of pure water ice at 150~K, and even the crystallization phase of ${\rm H_2O}$ at about 145~K, correspond to ${\rm CH_3NH_2}$ molecules desorbing from the surface of the water ice.
Our experimental results show that, when increasing slightly the exposure dose of methylamine on the np-ASW ice surface from 0.30~ML to 0.35~ML, the fraction of ${\rm CH_3NH_2}$ ice desorbing  from the  water ice surface at 137~K increases from 0.16~ML to 0.21~ML, while those desorbing directly from the HOPG surface at T$\geq$160~K decreases slightly from 0.16~ML to 0.14~ML, respectively. This means that by increasing the amount of methylamine on top of the water ice surface, ${\rm CH_3NH_2}$ molecules bind further with the water ice by hydrogen bonds and desorb between 120~K and 140~K, rather than diffuse through the water ice surface and desorb later from the HOPG substrate at 160-220~K. The fraction of the refractory ${\rm CH_3NH_2}$ molecules desorbing from the HOPG substrate after water ice sublimation is about 0.15~ML.

Our results are partially in agreement with those of \cite{2016Souda}, who observed a broad TPD curve of methylamine from the np-ASW ice of ${\rm D_2O}$ (8~ML), deposited  at 120~K on the nickel Ni(111) substrate. This latter comprises  two desorption peaks: one peak at 120~K corresponding to multilayer desorption of methylamine and another at about 150~K
corresponding to methylamine mixed with water coming from the aqueous solution. In addition, the TPD spectrum of \cite{2016Souda} shows no desorption peak of methylamine at surface temperature 160~K and above, suggesting that, in contrast to our case, no strong interaction between methylamine and the
underlying Ni(111) substrate takes place in their experiments. If our experiments performed on the np-ASW  ice of 10~ML thickness allowed the incorporation of methylamine from the water ice surface towards the graphite substrate, one can conclude that the diffusion of methylamine molecules through the non-porous water ice surface depends on the substrate below the water ice (HOPG, Ni). The nickel Ni(111) surface is expected to have fewer energetic binding sites than HOPG surface. However, if differences of substrates are seen for methylamine, one could say that similar differences are expected for formamide. The role of the substrate is discussed later.

\section{Analysis}\label{Analysis}

A simple mathematic model has been developed  to fit the TPD curves of formamide and methylamine in the sub-monolayer and monolayer regimes. At any surface
coverage, thermal desorption of molecules from the surface into the gas phase can be described in terms of an Arrhenius law, given by the Polanyi- Wigner equation (\ref{Eq2}) \citep{2012Noble,2001Dohnalek}
\begin{equation}\label{Eq2}
{\rm r(T)}=-\frac{\rm dN}{\rm dt}={\rm AN^ae^{-E_{des}/k_{B}T}},
\end{equation}

where ${\rm r(T)}$ is the desorption rate in (${\rm molecule\cdot cm^{-2}\cdot s^{-1}}$), N is the number density of molecules adsorbed on the surface, expressed in (${\rm molecules\cdot
cm^{-2}}$), a is the order of the desorption process, A is the pre-exponential factor expressed in (${\rm molecule^{-a+1}\cdot cm^{2a -2}\cdot s^{-1}}$), which can be considered to be the attempt frequency of molecules at overcoming the barrier to desorption, ${\rm k_B}$ is the Boltzmann constant
(${\rm k_B=1.38\times10^{-23}}$ J$\cdot$K$^{-1}$), t is the time in (s), ${\rm T}$ is the absolute temperature of the surface in (K), and ${\rm E_{des}}$ is the activation energy for desorption in (${\rm~kJ\cdot mol^{-1}}$) or in (K), where ${\rm~1~kJ\cdot mol^{-1}=120~K}$.

In the particular case of simple molecular adsorption, the pre-exponential  factor A may also be equated with the frequency of vibration of the bond between the molecule and substrate; this is because
every time this bond is stretched during the course of a vibrational cycle can be considered an attempt to break the bond and hence an attempt at desorption. However, we point out that this is a simplistic view of a complex problem. The physical observable parameter is the desorption flux; the pre-exponential factor and the binding energy are actually two degenerated parameters linked in the Polanyi-Wigner equation. Therefore, at anytime, a couple of solutions  (A, ${\rm E_{des}}$) should be considered.

Because the peaks of the TPD curves of formamide appear at a single temperature value for all the exposure doses up to 1.61~ML, the desorption of formamide is considered to be of first order, and
the desorption energy is expected to be dependent on the coverage \citep{2001Fraser}.

For TPD of first order, a=1, the desorption rate ${\rm r(T)}$ become
\begin{equation}\label{Eq3}
{\rm r(T)}=-\frac{\rm dN}{\rm dt}={\rm ANe^{-E_{des}/k_{B}T}},
\end{equation}
where the pre-exponential factor A for desorption is in (${\rm s^{-1}}$).
The desorption rate ${\rm r(T)}$ in (${\rm ML\cdot s^{-1}}$) can be expressed using  equation (\ref{Eq4})
\begin{equation}\label{Eq4}
{\rm r(T)_{\rm ML\cdot s^{-1}}}=\frac{\rm r(T)}{\rm {\beta^{-1}\rm (\int_{T_{min}}^{T_{max}}\frac{\rm dN}{\rm dt}.dT)}},
\end{equation}
where (${\rm \int_{T_{min}}^{T_{max}}\frac{\rm dN}{\rm dt}.dT})$ is the integrated area below the TPD curve of the designed 1~ML exposure dose between ${\rm T_{min}}$ and ${\rm T_{max}}$.

In this model, we consider that each exposure dose N in (ML) of molecules on the surface is the sum of the different fraction exposure doses ${\rm N_{i}}$ in (ML),
\begin{equation}\label{Eq9}
{\rm N=\sum_{i=1}^{n}{\rm N{_i}}=N_1+N_2+...},
\end{equation}
where ${\rm N_{i}}$ is the population of molecules desorbing from the surface with a desorption energy ${\rm E_{i}}$ at a surface temperature T.

We also assume that the desorption rate ${\rm r(T)}$ is the sum of the desorption rates ${\rm r_i(T)}$, i.e.
\begin{equation}\label{Eq7}
{\rm r(T)}=\sum_{i=1}^{n}{\rm  r_i(T)= AN_{1}e^{-E_{1}/k_{B}T}+AN_{i}e^{-E_{i}/k_{B}T}+...},
\end{equation}
where
\begin{equation}\label{Eq8}
{\rm r_i(T)}={\rm AN_{i}e^{-E_{i}/k_{B}T}}
\end{equation}
is the desorption rate of the molecular population,  ${\rm N_{i}}$, desorbing from the surface at a temperature T, with a pre-exponential factor A, and a desorption energy ${\rm E_{i}}$.

To fit the TPD data for an exposure dose N of molecules on the surface, we use the desorption rate  ${\rm r(T)}$, given by the equation~(\ref{Eq7}), then we set two values ${\rm
E_{min}}$ and ${\rm E_{max}}$ for the desorption energy of the molecules from the surface, and we choose a value for the pre-exponential factor A. When we run the program, we obtain the different
populations ${\rm N_{i}}$ of molecules desorbing from the surface with the desorption energy ${\rm E_{i}}$, at a surface temperature T, varying from ${\rm T_{min}}$  to ${\rm T_{max}}$. The best fit of the TPD data is obtained when the calculated curve matches well the experimental results, and the three computational parameters (A, ${\rm E_{i}}$, ${\rm N_{i}}$) are well constrained, where the
pre-exponential factor A is between ${\rm 10^{12}~s^{-1}}$ and ${\rm 10^{18}~s^{-1}}$,  the desorption energy barrier ${\rm E_{i}}$ is between ${\rm E_{min}}$  and ${\rm E_{max}}$,
and all the surface populations ${\rm N_{i}}$ satisfy the relation (\ref{Eq9}).

\subsection{Desorption energies of formamide}\label{w}

The top panel of Figure~\ref{Fig5} shows the best fits of the TPD curves of formamide for various surface doses (0.14~ML, 0.30~ML, 0.61~ML and 1.0~ML) on the graphite (HOPG) surface. These best fits are derived from the equation (\ref{Eq7}) with a pre-exponential factor ${\rm A=~10^{18}~s^{-1}}$. The use of a
low pre-exponential factor (${\rm A=~10^{12}~s^{-1}}$) in our modelling simulations does not reproduce the TPD curves on the graphite surface perfectly. The bottom panel of Figure~\ref{Fig5} shows
the surface population ${\rm N_{i}}$ of formamide molecules in (ML) as a function of the desorption energy ${\rm E_{i}}$ in (K) on the graphite (HOPG) surface for the first monolayer exposure coverage (${\rm N=1~ML}$). As shown in Fig. ~\ref{Fig5}, bottom panel, most of the surface population (${\rm \sum_{i=1}^{n}N_{i}=\sim 0.97~ML}$) of formamide on HOPG surface releases into the gas phase with a desorption energy distribution, ranging from  7460~K to 9380~K.

The same desorption energy distribution (7460-9380~K) is obtained by fitting the TPD data of formamide deposited on top of the np-ASW ice film of 10~ML thickness, covering the graphitic HOPG substrate. This main result is explained by the fast and efficient (>~95~$\%$) diffusion of ${\rm NH_2CHO}$ molecules from the  water ice surface towards the HOPG substrate, at  surface temperature, 40~K, or more probably during the warming-up phase of the ices.
{The bottom panel of Fig.~\ref{Fig5} also shows that for a pre-exponential factor ${\rm
A=~10^{18}~s^{-1}}$, the maximum of the desorption energy distribution of formamide  (7700~K) is higher than that of pure amorphous water ice (${\rm E_{des}=6490~K}$), calculated with the same pre-exponential value ${\rm A=~10^{18}~s^{-1}}$ from the parameters (${\rm E_{des}}$, A) of the amorphous  water ice  \citep{2001Fraser}. This means that  ${\rm NH_2CHO}$ molecules are more tightly physisorbed to the surface of the HOPG substrate than ${\rm H_2O}$ molecules.
\begin{figure*}
\centering
\includegraphics[width=18cm]{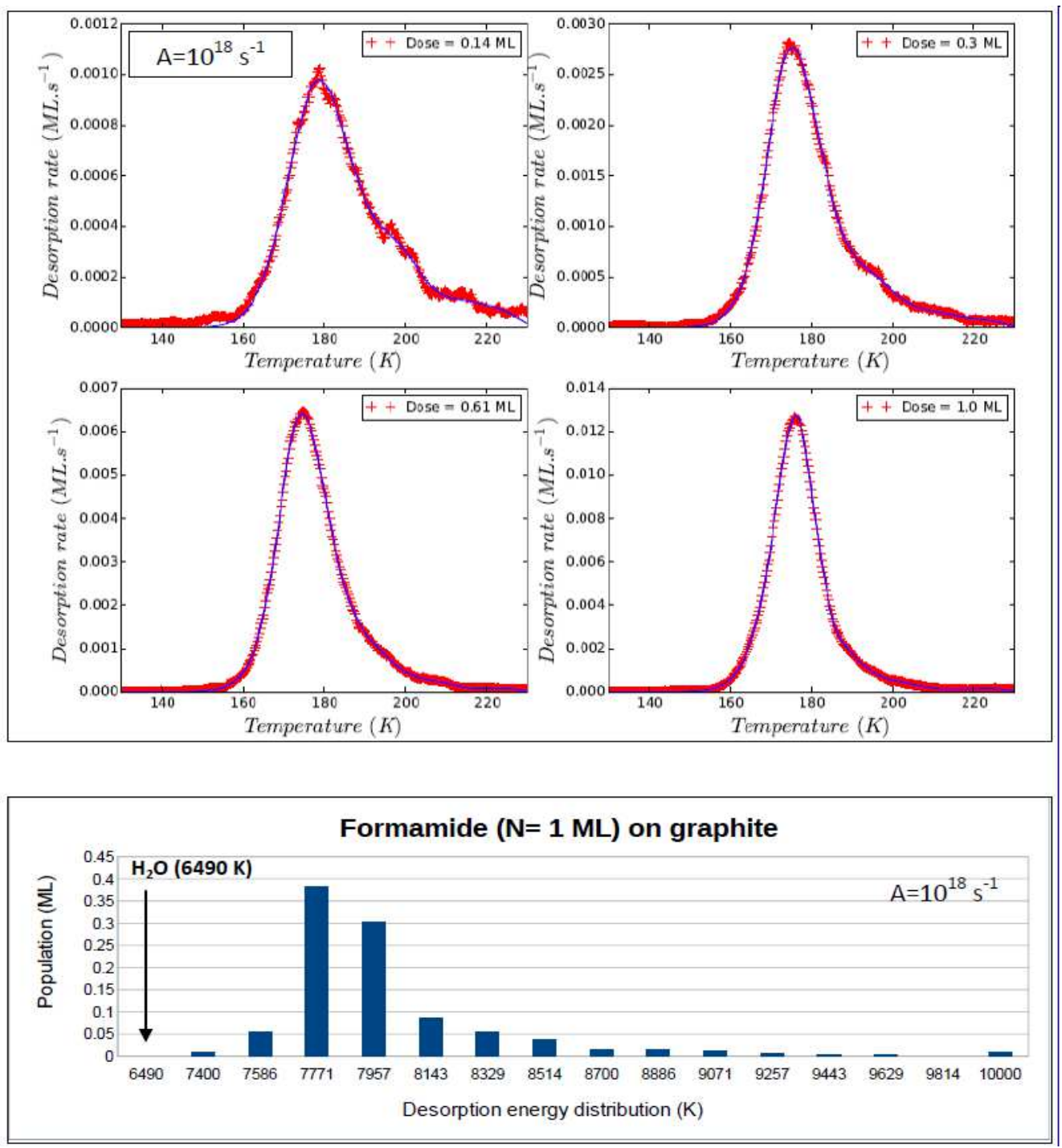}
\caption{Modelling results of the desorption rates in (${\rm ML.s^{-1}}$) and the desorption energy distribution in (K) of formamide on graphite (HOPG) surface.
Top panel: Red crosses indicate TPD data of ${\rm NH_2CHO}$ (m/z=45) on HOPG surface for different exposure doses of 0.14~ML, 0.30~ML, 0.61~ML, and 1.0~ML; Blue lines: The best fits of the TPD curves of ${\rm NH_2CHO}$ from the HOPG surface, calculated with the pre-exponential factor ${\rm A=10^{18}~s^{-1}}$ are shown. Bottom panel: The surface population in (ML) of ${\rm NH_2CHO}$ on HOPG  surface as a function of the desorption energy in (K) is shown. The desorption energy distribution (${\rm E_{i}}$) and the surface population (${\rm N_{i}}$) of formamide are derived from equation (\ref{Eq7}) with the best pre-exponential factor ${\rm A=10^{18}~s^{-1}}$ for the exposure dose ${\rm N=\sum_{i=1}^{n}N_{i}=1~ML}$. } \label{Fig5}
\end{figure*}

\subsection{Desorption energies of methylamine }\label{methylamine1}

Similar calculations of the desorption energy distribution of methylamine have been carried out both on graphite HOPG and np-ASW ice surfaces.
Assuming a first order desorption process of methylamine on graphite surface, the best fits of the TPD curves for various values of the surface exposure coverage, N (0.25~ML, 0.46~ML,
0.71~ML, and 1.0~ML) of methylamine, obtained from the model with the pre-exponential factor ${\rm A=~10^{12}~s^{-1}}$, are shown in Figure~\ref{Fig7}.

\begin{figure*}
\centering
\includegraphics[width=18cm]{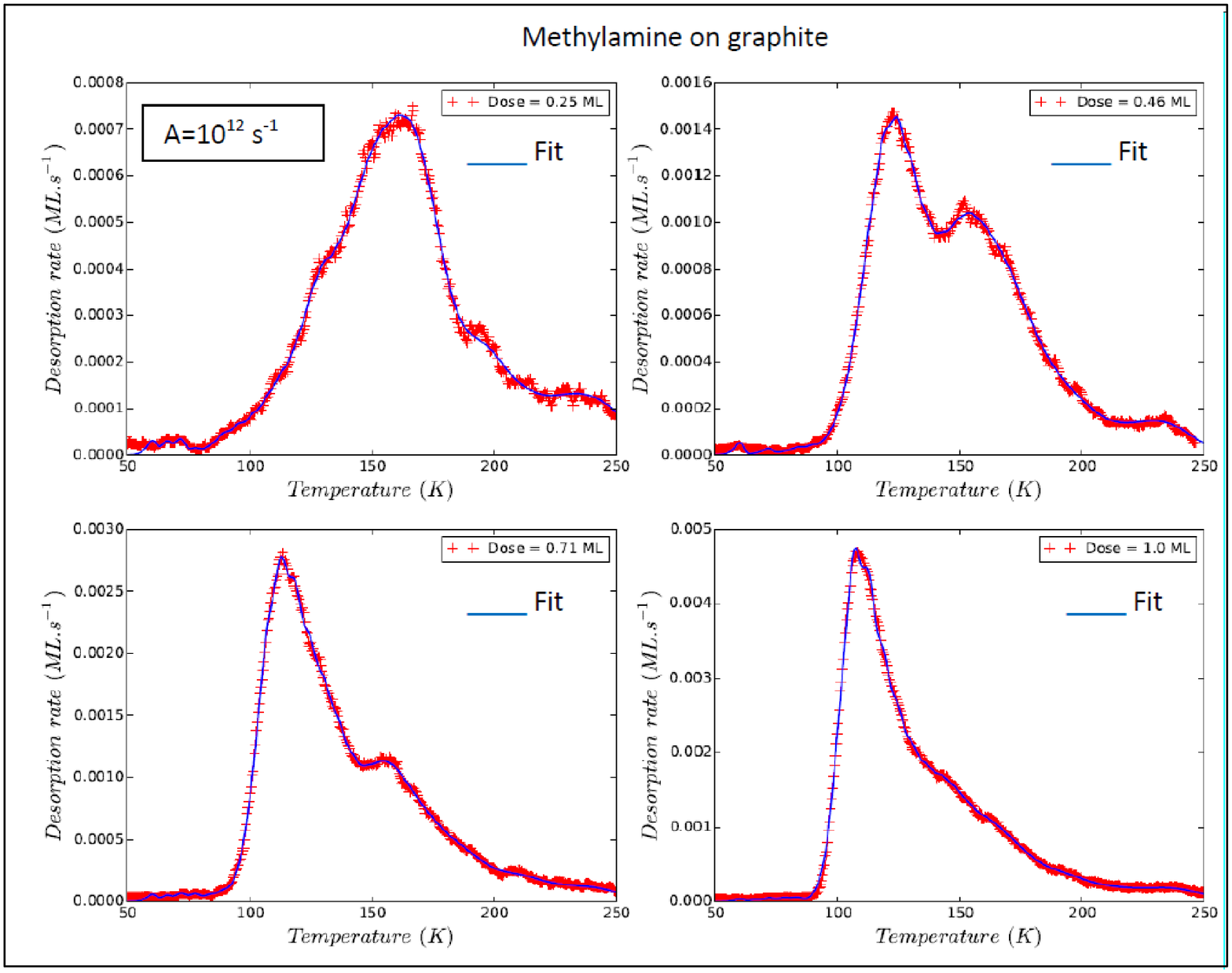}
\caption{Modelling results of the desorption rates in (${\rm ML.s^{-1}}$) of methylamine on graphite (HOPG) surface.
Red crosses: TPD data of ${\rm CH_3NH_2}$ (m/z=31)  on HOPG surface for different exposure doses of 0.25~ML, 0.46~ML, 0.71~ML, and 1.0~ML are shown; Blue lines: The best fits of the TPD curves of ${\rm CH_3NH_2}$ from the HOPG surface, calculated with the best pre-exponential factor ${\rm A=10^{12}~s^{-1}}$ are represented.}\label{Fig7}
\end{figure*}

The top panel of Fig.~\ref{Fig6} shows the surface population ${\rm N_{i}}$ in (ML) of solid methylamine on graphite (HOPG) surface as a function of the desorption energy ${\rm E_{i}}$ of methylamine, in (K), for the first monolayer exposure coverage (${\rm N=1~ML}$).
The desorption energy distribution of methylamine from graphite is found to range between 3010~K and 8420~K, where a large percentage ($\sim$~0.80~ML) of molecules desorbing  with low binding energies (3010-4454~K) correspond to the multi-layer and monolayer desorptions of ${\rm CH_3NH_2}$ species from the graphite surface. The range of the desorption energy (5050-8420~K) is assigned to the surface population ($\sim$~0.15~ML) of methylamine that desorbs from the very energetic adsorption sites of the graphite (HOPG) surface.

In the case of Figure~\ref{Fig6}, bottom panel, where an exposure dose N=0.30~ML of solid methylamine is deposited on top of the np-ASW ice film, the desorption energy distribution of these molecules is reduced to the range (3900-8420~K), with traces of small energies (2770-3610~K) corresponding to methylamine multi-layer desorption.
The bottom panel of Fig.~\ref{Fig6} also shows that a surface population of about 0.15~ML of ${\rm CH_3NH_2}$ leaves the water ice surface with desorption energies (3900-4500~K), which is lower than the binding energy value (4930 ~K) of crystalline water ice, calculated with the same pre-exponential factor ${\rm A=~10^{12}~s^{-1}}$ from the desorption parameters (${\rm E_{des}}$, A) of \cite{2001Fraser}. Furthermore, the desorption energies (5050-8420~K), which are higher than those of the amorphous water ice, are attributed to the fraction (${\rm \sum_{i=1}^{n}N_{i}=\sim 0.15~ML}$) of methylamine population that diffuses through the water ice surface towards the highest binding sites of the HOPG substrate, and then desorbs into the gas phase after water ice sublimation.

\begin{figure*}
\centering
\includegraphics[width=18cm]{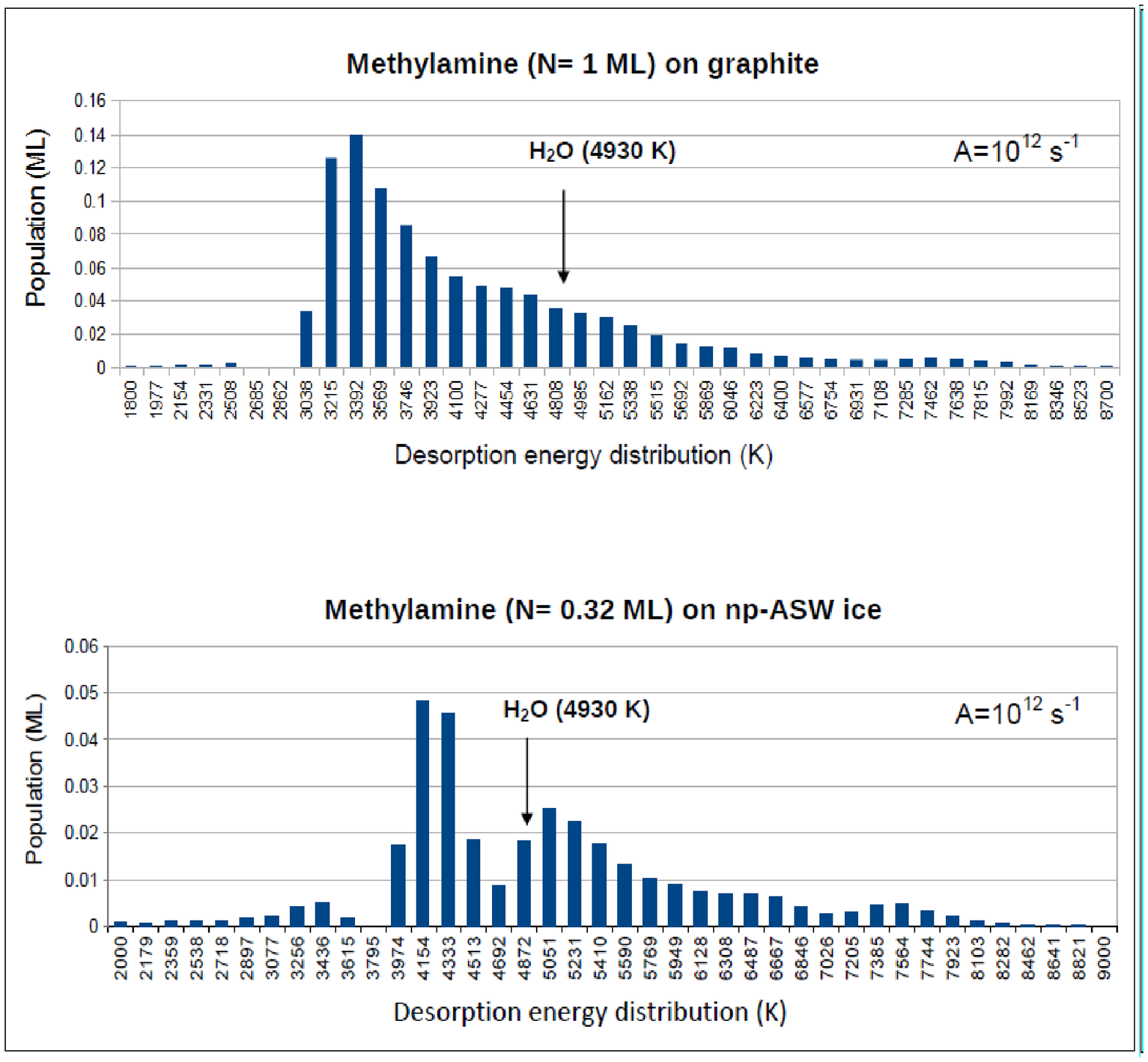}
\caption{Modelling results of the desorption energy distributions in (K) of methylamine on graphite (HOPG) and np-ASW ice surfaces.
Top panel: Surface population  in (ML) of ${\rm CH_3NH_2}$ (m/z=31) as a function of the desorption energy in (K) for an exposure dose N=1~ML of ${\rm CH_3NH_2}$ on the HOPG surface. Bottom panel: Surface population in (ML) of ${\rm NH_2CH_3}$ as a function of the desorption energy in (K)  for an exposure dose N=0.30~ML of ${\rm CH_3NH_2}$ on the surface of the np-ASW ice film of 10~ML thickness, prepared at 110~K and cooled down to 40~K. The desorption energy distribution (${\rm E_{i}}$) and the surface population (${\rm N_{i}}$) of methylamine are derived from equation (\ref{Eq7}) with the best pre-exponential factor ${\rm
A=10^{12}~s^{-1}}$ for the two exposure doses ${\rm N=\sum_{i=1}^{n}N_{i}=1~ML}$ on HOPG and ${\rm N=\sum_{i=1}^{n}N_{i}=0.30~ML}$ on np-ASW ice surfaces.}\label{Fig6}
\end{figure*}

\section{Discussion}\label{Discussion}

We developed a model using the Polanyi-Wigner equation to reproduce the TPD profiles and determine the desorption energy distributions of ${\rm NH_2CHO}$ and  ${\rm CH_3NH_2}$
molecules on graphite (HOPG) and non-porous ASW ice surfaces in the sub-monolayer and monolayer regimes.
The desorption energy distributions, ${\rm E_{des, dist}}$, of formamide and methylamine  are obtained with different pre-exponential factors: A=10$^{18}$~s$^{-1}$ for formamide and A=10$^{12}$ s$^{-1}$ for methylamine. In the case of formamide (${\rm NH_2CHO}$), the decrease in the pre-exponential factor A from 10$^{18}$~s$^{-1}$ to 10$^{12}$ s$^{-1}$, affects the quality of the TPD fits, and reduces the value of the desorption energy of the molecules by 30~$\%$.
As already mentioned, A and E$_{\rm {des}}$ are degenerated parameters that can be derived from a desorption flux at a given temperature T. However, the dynamic of desorption gives more constraints on the couple of parameters. Experimentalists seeks for getting the best fit from their experiments, and as evidence, the best fit is obtained with extreme differences in the pre-exponential factor  values (A= 10$^{18}$ and A= 10$^{12}$ s$^{-1}$) for two apparently similar molecules. The degeneracy of A and E${\rm _{des}}$  has been already estimated and discussed previously. Simple molecules (like H$_2$) seems to have a pre-exponential factor for desorption usually close to 10$^{12-13}$ s$^{-1}$ \citep{2006Amiaud}. On the contrary, \cite{2015Dronin} have shown that the same molecule (CH$_3$OH) could have different pre-exponential factors for different substrates or regimes (multi-layers or sub-monolayer). These authors proposed a method to derive the pre-exponential factor from TPD data.
It has been suggested by \cite{2005Tait} that the pre-exponential factor A for the desorption depends on the potential interaction of the adsorbate substrate system, and the number of degrees of freedom of the molecule when it passes from the adsorbate state to the gas-phase transition state. These authors have demonstrated the increase of the  pre-factor A for desorption with the chain length of the molecule from the value 10$^{13}$~s$^{-1}$ to 10$^{19}$~s$^{-1}$, and have explained this variation of A  by the increase in the rotational entropy available to the molecules in the gas-like transition state for desorption.
The difference in the pre-exponential factor values is an experimental fact. In this work, we want to determine the binding energy distributions of molecules, and once done,  we need to be able to compare the different values, originating from different works or analysis methods.

Table~\ref{table1} shows the maximum of the desorption energies (${\rm E_{des,max}}$) of formamide and methylamine from different substrates, calculated from the model, with the original suitable pre-exponential factors (A), and with a fixed value (${\rm A=10^{12}~s^{-1}}$) for better comparison with other relevant astrophysical molecules, using the following desorbing flux relationship,
\begin{equation}\label{Eq11}
{\rm r(T)}={\rm A_1e^{-E_{1, des}/k_{B}T}}={\rm A_2e^{-E_{2, des}/k_{B}T}},
\end{equation}

where ${\rm A_1}$ and ${\rm A_2}$ are the pre-exponential factors in (s$^{-1}$), associated with the desorption energies ${\rm E_{1, des}}$ and ${\rm E_{2, des}}$, respectively, of the adsorbed molecules on the surface, and T is the experimental  temperature in (K) of the surface, at which the desorption is observed.

In the Table~\ref{table1}, we can compare the value of the binding energy of methanol  (${\rm CH_3OH}$) on graphite with those of methylamine and formamide. We can see that even though the pre-exponential factors and analysis methods are different, all the approximated values of binding energies given here are close and can be compared to  the value of the binding energy of the amorphous water ice derived from the same calculation method. The important result here is that, in any case, the binding energy of CH$_3$OH  is lower than that of water.
Table~\ref{table1} also shows that for the same employed pre-exponential factor ${\rm A=10^{12}~s^{-1}}$, the maximum of the desorption energy distribution (${\rm E_{des, max}=5265~K}$) of ${\rm NH_2CHO}$ from graphite (HOPG) substrate is smaller than the value (6871~K) of the mean desorption energy obtained from silicate (${\rm SiO_2}$) interstellar grain analogue surface by \cite{2014Dawley}. This difference between our result and that of \cite{2014Dawley} is an indication that the desorption energy of formamide depends on the substrate itself.

More precisely, we find that the binding energy of 0.1~ML of formamide adsorbed on graphite substrate is 5600-5900~K, calculated with A=10$^{12}$~s$^{-1}$. These values are smaller than the value given on silicate substrate \citep{2014Dawley}. This difference may not be only related to the composition of the substrate (silicate, graphite), but could also be due to the difference in the morphology of the substrate, which can easily change the binding energy distribution of the molecules by a few tens of percent \citep{2009Fillion}. In our experiments, the desorption energy of formamide from the surface of the non-porous ASW ice cannot be measured because the water substrate desorbs before formamide, that is deposited on top of it. Actually, the whole binding energy distribution of formamide from graphite (5056-6990~K), calculated with A=10$^{12}$~s$^{-1}$ for 1~ML coverage, is higher than the binding energy values of water ice in its amorphous (4810~K), and crystalline state (4930~K).
Unlike formamide, solid methylamine can desorb under different forms, in the multi-layer regime (${\rm NH_2CH_3-NH_2CH_3}$) with low desorption energies (3010-3490~K) from the np-ASW ice surface with desorption energies (3900-4500~K) and even later from the HOPG substrate with higher desorption energies (5050-8420~K). The different desorption energies of methylamine calculated in this work with respect to that of pure amorphous water ice (4810~K) is due to its wetting ability compared to ${\rm H_2O}$ molecules. In fact, ${\rm CH_3NH_2}$ species are able to spread readily and uniformly over the surface of the graphite (HOPG) or the amorphous water ice by hydrogen bonds, thereby forming a thin and continuous film in which physisorbed molecules populate a large distribution of surface binding sites.
The proportion of molecules desorbing, prior, during, or after the water ice varies with the doses of methylamine, but each time the fraction of methylamine desorbing after the water ice is about 0.15~ML. However, the fraction of methylamine co-desorbing with the crystalline phase of  the water ice is very small at about 0.03~ML. We can see in Figure~\ref{Fig4} that the desorption  of methylamine is very reduced during the desorption of the water ice at around 150~K. The desorption of methylamine at 137~K, before the sublimation of the water ice, is delayed compared with its desorption without water at T=106~K (Figure~\ref{Fig4}), indicating a stronger interaction of methylamine with the water molecule than with itself.
In most astrophysical models, the whole desorption of each molecule is described by its (unique) binding energy. For methylamine, the main question is which binding energy shall we propose. Usually, binding energies are measured for pure species and therefore, the answer would be E${\rm _{des}}\simeq$3000~K  with A=10$^{12}~$s$^{-1}$. But in an astrophysical context,  the proportion of methylamine in icy mantles is certainly less than a few $\%$ as it is not positively detected in ices \citep{2015Boogert}. Therefore, it would be better to take the low concentration limit of our experiments ($\sim$0.15~ML), which would lead to an estimation of the binding energy at around 6000~K, well above the value of those of the water ice. This low coverage value is certainly very dependent on the substrate and represents the wetting properties of the surface for a specific molecule. This is probably why \cite{2016Souda} did not find a late desorption of methylamine in the case of metallic surface Ni (111).
The key question is to find out whether at low concentration, the binding energy of a specific molecule is higher than the binding energy of the water ice. Table~\ref{table1} reveals that the binding energies of methanol (CH$_3$OH) and formaldehyde (H$_2$CO) are lower than that of amorphous (H$_2$O) water ice, and so we can consider these molecules as volatile, or at least they would desorb with the water ice mantle. For some other molecules, such as NH$_2$OH and NH$_2$CHO, the binding energy is higher than that of (H$_2$O) water ice, and we can consider these molecules as refractory species, compared to water molecules. Refractory does not have the same meaning here as sometimes used in shock-regions studies or in geology. Here, it is the remaining part of the molecular mantle after the sublimation of water.
Before this study, we were wondering if some simple chemical properties could explain the common trends of the binding energies. It is usually discussed in terms of hydrophilic and hydrophobic proprieties of the molecules or in terms of hydrogen bonding formation with water.  We can sort molecules in two groups: those with a binding energy less than that of water, such as CH$_3$OH, NH$_3$, H$_2$CO, and another group with a binding energy, at low coverage, that are higher than the binding energy of water, among which we can find  H$_2$O$_2$ \citep{2017Dulieu}, NH$_2$OH \citep{2012Congiu}, and NH$_2$CHO.
As a matter of fact, the hydrophobic nature of the methyl ( -CH$_3$) group, or the hydrophilic nature of the amino (-NH$_2$) or hydroxyl (-OH) groups, cannot explain our observations. Only detailed calculations could probably give a real understanding of this empirical sorting.

Formamide and methylamine that diffuse completely or partly through the ASW ice surface towards the graphitic substrate and desorb into the gas phase at higher temperatures ($\geq$160~K)
are considered as refractory species, which can remain frozen on dust grains even after water ice sublimation. This result leads us to consider different scenarios, in which formamide and methylamine can be formed either in the gas phase and then accreted onto the ice surface or formed by grain-surface chemistry. In the case of the gas-phase formation, the diffusion of the adsorbed
molecules on the colder ice mantle at 10-20~K is an open question, since it depends on the thickness, porosity, and chemical composition of the ice mantle~\citep{2013Mispelaer,2015Lauck}.
In the case of grain-surface chemistry,  ${\rm NH_2CHO}$ and ${\rm CH_3NH_2}$ molecules may be formed on icy grain mantles in the interstellar medium, by exothermic reactions between mobile
radicals, generated by ultraviolet (UV) radiation, or trapped within the bulk of ice mantles of dust grains \citep{2012Taquet}, probably at 30-40~K, once the protostar starts heating its immediate
surroundings, as has been suggested by grain surface chemistry models of \cite{2006Garrod}. The diffusion of the molecules within the water ice to the surface of dust grains by hydrogen bonding
formation depends on the morphology of the ices \citep{2015Lauck} and  mainly on the surface of the dust grain below the water ice. The diffusion of ${\rm NH_2CHO}$ molecule from the water ice surface to the graphitic substrate is possibly favoured by the formation of larger groups of segregated ${\rm NH_2CHO}$ molecules, which bind together by intermolecular van der Waals interactions, and with ${\rm H_2O}$ molecules by hydrogen bonds through the amino (${\rm -NH_2}$) and  carbonyl (-C=O) functional groups. The fast diffusion of ${\rm NH_2CHO}$ molecules through the water ice may occur during the reorganization of the ices, and the transition phase from the amorphous to the crystalline structure of the ASW ice at about 145~K \citep{2013Mispelaer, 2014Dawley}.
 In the interstellar medium, dust grains can be composed of
relevant astrophysical materials, such  as amorphous carbon (a-C), amorphous silicon (a-Si), and amorphous silicates (${\rm Mg_2SiO_4}$ and ${\rm Fe_2SiO_4}$). Amorphous carbon (a-C), such
as a diamond, has been shown to consist of graphitic islands, composed of compact polycyclic aromatic ring structures, poorly connected with each other \cite{1989Duley}. These astrophysical relevant surfaces, characterized by deepest binding sites and highest binding energies, in comparison to the surface of the amorphous water ices, may affect the diffusion proprieties of the molecules. So
methylamine and formamide molecules interact strongly with these amorphous carbon and silicon surfaces after water ice evapouration. Formamide molecules are expected to fully diffuse from the water ice surface to these interstellar analogue surfaces.

If dust grains are heated, as in the case of those located nearby a newly born protostar, the external heat provided by the protostars increases the surface temperature of
the ice mantle, provoking desorption from the ice back to the gas phase of the volatile molecules, such as CO, ${\rm CO_2}$, and ${\rm H_2O}$ or CH$_3$OH. As the thickness of the water ice on the dust grain is reduced, the non-volatile formamide molecules diffuses  through the water ice and populates the highest bindings sites of dust grains. Therefore the presence in the gas phase of refractory molecules, if these molecules come from the solid phase, would require a higher final temperature of dust grains. These refractory molecules should be observed on a separate snow line. Furthermore, this thermal desorption should be accompanied by a blending of the species and some induced reactivity, as described in for example \cite{2013Theule, 2017Dulieu}, and in \cite{2016Souda} for the H-D exchange in ices.
But the presence of molecules in the gas phase coming from the solid phase can arise from the sputtering of the grain mantle, especially in shocks (e.g. \cite{1983Draine,1994Jones,1994Tielens}). This type of desorption process is not at thermal equilibrium, but is induced by gaseous bombardment with an energy greater than the surface binding energies, provoking the ejection of the material in gas environments. This kind of ballistic kick out mechanism should be sensitive to the binding energy of the molecules.

Recent astronomical observations of \cite{2012bacmann} have suggested that complex organic molecules (COMs) are actively  formed in prestellar cores in L1689B before the gravitational collapsing phase leading to
protostar formation and even before the warm-up phase of the grains. Many COMs have been detected in cold dark prestellar cores  at low  temperature (T$\thicksim$10 K). This includes dimethyl ether (${\rm CH_3OCH_3}$), methyl formate (${\rm CH_3OCHO}$), acetaldehyde (${\rm CH_3CHO}$), ketene (${\rm CH_2CO}$),  cyclopropenone (${\rm c-C_3H_2O}$), propynal (${\rm HCCCHO}$), vinyl cyanide (${\rm CH_2CHCN}$), and propyne  (${\rm CH_3CCH}$) \citep{2016Jimenez,2014Vastel, 2012bacmann}.
Here one of the main efficient mechanism included in astrochemical models to feed back the gas phase is the impulsive heating of dust grains by X-rays and cosmic rays, already proposed earlier by \cite{1985leger}. Here again, the binding energy of the molecules is still a selective parameter because it explicitly enters in the calculation of the desorption of each type of molecules. Although this is not exactly the same mechanism than found in thermal desorption, in practice, models sort molecules depending on their binding energies. Therefore, in the context of prestellar cores, a very high binding energy also prevents any return in the gas phase. The second efficient mechanism for populating the gas phase is the chemical desorption (or reactive desorption)  \citep{2000Takahashi, 2007Garrod, 2013Dulieu}. Here, the energy released during the formation of the molecule itself is the promoter of the desorption. But the probability of desorption has been proposed to decrease exponentially with the binding energy, among others factors \citep{2016Minissale}.
The COMs detected in cold dark prestellar cores regions are likely to originate  from gas-phase reactions at 10-20~K of precursor molecules, such as formaldehyde and methanol, which are previously formed on icy grain surfaces, then ejected into the gas phase by desorption induced by chemical reaction, and finally followed by gas-phase chemistry via ion molecules \citep{2013Vasyunin}, neutral-neutral, and radiative association reactions \citep{2017Vasyunin}. These molecules detectable in the gas phase can in turn be involved in  gas-phase reactions to form more complex organic molecules. It is therefore, interesting to explore the interaction of these detected COMs with the ASW ice and with various astrophysical relevant substrates.
Formamide and methylamine are probably forming in regions where CO has also started to freeze-out, i.e. in dense molecular clouds. In these regions, the icy mantle might be composed of water
ice in the presence of CO molecules or a mixture of ${\rm CO_2}$ and ${\rm H_2O}$ ices. However, since CO desorbs at 30-50~K from the amorphous water ice surface \citep{2012Noble}, it is probably that } its presence with ${\rm H_2O}$ molecules does not affect the diffusion of formamide and methylamine from the water ice surface towards the substrate, such as graphite HOPG, at the expected temperatures of 110-140~K. Nevertheless, it would be interesting  to perform further adsorption-desorption  laboratory experiments of formamide and methylamine on mixed or layered  ${\rm CO}$, ${\rm CO_2}$, and H$_2$O ices  to understand  the effect of this relevant
astrophysical molecule on  the diffusion process of the complex organic molecules through the ASW ice. It would also be interesting to explore other substrates, such as amorphous silicon (a-Si) and amorphous carbon (a-C), which better represent the interstellar dust grains composition than HOPG to study the effect of the substrate on the desorption energies and the diffusion of these two molecules, and other COMs from the water ice surfaces.

\begin{table*}
\centering \caption{Desorption energy distributions (${\rm E_{des, dist}}$) in (K), column 3, of  ${\rm NH_2CHO}$ and ${\rm CH_3NH_2}$ obtained in this work with the best-fit pre-exponential factor A values, expressed in (s$^{-1}$). Column 4 gives the maximum of the desorption energy (${\rm E_{des-max}}$), in (K) obtained with the best-fit pre-exponential factor A values, in (s$^{-1}$) for ${\rm NH_2CHO}$, ${\rm NH_2CH_3}$, and other astrophysical relevant molecules  (${\rm NH_2OH}$, ${\rm CH_3OH}$, ${\rm H_2CO}$, and ${\rm H_2O}$) taken from the literature. For a comparison,  column 5 gives the desorption energies  (${\rm E_{des, max}}$) of all the molecules calculated with the typical pre-exponential factor A=10$^{12}$ s$^{-1}$ for various grain surfaces, using equation~(\ref{Eq11}) of flux desorption (${\rm A_1e^{-E_{1, des}/k_{B}T}}$=${\rm A_2e^{-E_{2, des}/k_{B}T}}$), and the values of A and ${\rm E_{des, max}}$ in column 4}. \label{table1}
\begin{tabular}{c|c|c|c|c|c}
\hline\hline

Molecule                   & Substrate         &    ${\rm E_{des, dist}}$; (A)             & ${\rm E_{des, max}}$; (A)                        &    ${\rm E_{des, max}}$    &  References \\
                                 &                          &                                                                    &                                                                  &   (A=10$^{12}$ s$^{-1}$)   &                        \\
\hline
                                &                          & ${\rm K}$; (s$^{-1}$)                                    &   ${\rm K}$; (s$^{-1}$)                            & ${\rm K}$                         &                          \\
\hline
${\rm NH_2CHO}$  & HOPG              & $7460-9380$; (10$^{18}$)                            &     7700; (10$^{18}$)                              & 5265                                 &     This work\\
                                 &                          &                                                                      &                                                                &  ${\rm E_{des, dist}}$=5056-6990  &              \\
${\rm NH_2CHO}$  & H$_2$O           &                                                                       &                                                               &                                         &     This work \\
${\rm NH_2CHO}$  & ${\rm SiO_2}$ &                                                                       &     7397; (10$^{13}$)                              &  6871                                 &   \cite{2014Dawley} \\
\hline
${\rm CH_3NH_2}$  & ${\rm NH_2CH_3}$ &   $3010-3490$; (10$^{12}$)                &    3307; (10$^{12}$)                                &  3307                                  &   This work \\
${\rm CH_3NH_2}$  & H$_2$O                  &    $3900-4500$; (10$^{12}$)                &  4269; (10$^{12}$)                                  &  4269                                  &   This work \\
${\rm CH_3NH_2}$  & HOPG                     &   $5050-8420$; (10$^{12}$)                &  5100; (10$^{12}$)                                  &  5100                                  &   This work \\
\hline
${\rm NH_2OH}$    & Silicate                     &                                                              & 6518; (10$^{13}$)                                    &   6080                                   &   \cite{2012Congiu} \\
\hline
 ${\rm CH_3OH}$    & ${\rm CH_3OH}$    &                                                            &  4989;  (5$\times$10$^{14}$)                    &   4169                                     &  \cite{2015Dronin} \\
 ${\rm CH_3OH}$    & ${\rm H_2O}$          &                                                          & 4366; (2$\times$10$^{12}$)                       &     4310                                   & \cite{2014Domenech} \\
 ${\rm CH_3OH}$    & ${\rm graphite}$       &                                                          & 5454; (10$^{16}$)                                      &  4257                                       &  \cite{2015Dronin} \\
 \hline
 ${\rm H_2CO}$     &  ${\rm H_2O}$       &                                                              &  3247; (10$^{13}$)                                     &     2899                                    &    \cite{2012Noble} \\
 ${\rm H_2CO}$     &  Silicate                 &                                                               &  3729; (10$^{13}$)                                    &     3040                                     & \cite{2012Noble} \\
 \hline
 ${\rm H_2O}$       &                                        &                                                              &                                                                   &                                                  &                              \\
 Amorphous          &    Au                             &                                                              &5600; (10$^{15}$)       &    4810                                 &     \cite{2001Fraser}\\
 Crystalline           &    Au                            &                                                              & 5773; (10$^{15}$)        &   4930                               &  \cite{2001Fraser} \\
  \hline\hline
\end{tabular}
\end{table*}

\section{Conclusions}\label{conclusions}

The desorption of formamide ${\rm NH_2CHO}$ and methylamine ${\rm CH_3NH_2}$ molecules has been investigated experimentally on graphite (HOPG) and np-ASW water ice surfaces in the sub-monolayer and monolayer regimes, using the temperature programmed desorption technique (TPD).

Experimental results show an efficient diffusion of more than 95 $\%$ of formamide through the np-ASW ice film of 10~ML thickness towards the graphite (HOPG) substrate. The large percentage of solid formamide  ($>$ 0.95~ML) bound to the graphitic substrate desorbs at higher surface temperature, 176~K, whatever the surface coverage, after the desorption of the water ice at about 150~K. The diffusion of ${\rm NH_2CHO}$ is likely to occur during the warming-up phase of the ices at a short timescale of few seconds.

${\rm CH_3NH_2}$ molecules physisorbed on the ASW ice desorbs at 137~K, before the desorption of the ${\rm H_2O}$ molecules at 150~K, and even  from the energetic sites of the HOPG surface at higher temperatures 160-220~K. Because of the wetting ability of the methylamine compared to water, the fraction of solid methylamine that diffuses through the water ice surface towards the graphitic substrate is about 0.15~ML.
The amounts of formamide and methylamine desorbing from the graphite dust grains after water ice sublimation are considered as refractory species, which can enrich the gas phase of warm interstellar  environments.

We analysed the binding energy distributions and proposed a convenient way to compare the binding energies directly, by comparing these energies with an arbitrary fixed pre-exponential factor (A=10$^{12}$ s$^{-1}$). The main difference in behaviour of these two molecules is probably because the entire binding energy distribution of formamide  (${\rm E_{des}=}$ ${\rm 5056-6990~K}$) is higher than the value of the amorphous water desorption (${\rm E_{des}=4810~K}$). On the contrary, the binding energy distribution of methylamine  (${\rm E_{des}=3010-8420~K}$) is distributed over this value.
There is no obvious link between the chemical functional groups (amino, hydroxyl, methyl, and carbonyl) and the binding energy distribution on our graphite template. The simple idea that all the molecules should desorb during the sublimation of the ice mantle is certainly unrealistic and illusive. The difference in the binding energies of formamide and methylamine could have an impact on the composition of the gas-phase environments, in particular in the comae of comets, but also in hot cores, hot corinos, and protoplanetary disks.

\begin{acknowledgements}

This work was supported by the Programme National ''Physique et Chimie du Milieu Interstellaire  (PCMI)'' of CNRS / INSU with INC / INP co-funded by CEA and CNES. H.C would like to thank Pr Darek. Lis, Pr Jean-Hugues Fillion of the LERMA team for their support and helpful corrections. The authors acknowledge the anonymous referee for constructive comments that greatly improved the paper. T.N is granted by the LabEx MICHEM. Formamide is a target molecule of SOLIS, a NOEMA key programme led by C. Ceccarelli and P. Caselli.

\end{acknowledgements}
\bibliographystyle{aa}       
\bibliography{REF.formmidemethyamine}

\end{document}